\documentclass[cmp]{svjour}

\pdfoutput=1

\usepackage{amsmath}
\usepackage{amsfonts}
\usepackage{amssymb}
\usepackage{graphicx}%
\begin{document}

\title{The Makeenko--Migdal equation for Yang--Mills theory on compact surfaces}

\title{The Makeenko--Migdal equation for Yang--Mills theory on compact surfaces}
\author{Bruce K. Driver\inst{1} \and Franck Gabriel\inst{2} \thanks{Supported by ERC grant,
\textquotedblleft Behaviour near criticality,\textquotedblright\ held by M. Hairer} \and
Brian C. Hall\inst{3} \thanks{Supported in part by NSF grant DMS-1301534} \and
Todd Kemp\inst{1} \thanks{Supported in part by NSF CAREER award DMS-1254807}}

\institute{University of California San Diego, Department of Mathematics, La Jolla, CA 92093 USA \and
Mathematics Institute, Zeeman Building, University of Warwick, Coventry CV4 7AL UK \and
University of Notre Dame, Department of Mathematics, Notre Dame, IN 46556 USA}

\maketitle

\begin{abstract}
We prove the Makeenko--Migdal equation for two-dimensional Euclidean
Yang--Mills theory on an arbitrary compact surface, possibly with boundary. In
particular, we show that two of the proofs given by the first, third, and
fourth authors for the plane case extend essentially without change to compact surfaces.

\end{abstract}

\email{bdriver@ucsd.edu \and F.Gabriel@warwick.ac.uk \and bhall@nd.edu \and tkemp@math.ucsd.edu}

\section{Introduction}

The Euclidean Yang--Mills field theory on a surface $\Sigma$ describes a
random connection on a principal bundle over $\Sigma$ for a compact Lie group
$K,$ known as the structure group. Work of A. Sengupta \cite{Sen93,Sen97,Sen3,Sen4}
gave a formula for the expectation value of any gauge-invariant function
defined in terms of parallel transport along the edges of a graph $\mathbb{G}$
in $\Sigma.$ (Related work was done by D. Fine \cite{Fine1,Fine2} and E.
Witten \cite{Witten1,Witten2}.) This theory was then further developed
\cite{LevySurf} and generalized \cite{LevyMarkov} in the work of T. L\'{e}vy.
Sengupta's formula (generalizing Driver's formula \cite[Theorem 6.4]{Dr} in
the plane case) is given in terms of the \textit{heat kernel} on the group
$K.$ (See Section \ref{YMsurf.sec}.) One noteworthy feature of the formula is
its invariance under area-preserving diffeomorphisms of $\Sigma.$

The typical objects of study in the theory are the Wilson loop functionals,
given by%
\begin{equation}
\mathbb{E}\{\mathrm{trace}(\mathrm{hol}(L))\}, \label{wilsonLoop}%
\end{equation}
where $\mathbb{E}$ denotes the expectation value with respect to the
Yang--Mills measure, $\mathrm{hol}(L)$ denotes the holonomy of the connection
around a loop $L$ traced out in a graph $\mathbb{G},$ and the trace is taken
in some fixed representation of $K.$ The diffeomorphism-invariance of the
theory is reflected in Sengupta's formula: the expectation \eqref{wilsonLoop}
is given as a function (determined by the topology of the graph and of
$\Sigma$) of all the areas of the faces of $\mathbb{G}.$

A key identity for calculating Wilson loops is the \textit{Makeenko--Migdal equation}
\cite[Equation 3]{MM} for Yang--Mills theory.  For the plane case, 
%Meanwhile, in \cite{LevyMaster}, L\'{e}vy established a rigorous version of
%the \textit{Makeenko--Migdal equation} \cite[Equation 3]{MM} for Yang--Mills
%theory on the plane. For Yang--Mills theory in the plane,
the Makeenko--Migdal equation takes the form \eqref{MMun} below,
as worked out by V. A. Kazakov and I. K. Kostov in \cite[Equation 24]{KK} (see also 
\cite[Equation 9]{KazakovU(N)} and \cite[Equation 6.4]{GG}. We take $K=U(N)$ and
we use the bi-invariant metric on $U(N)$ whose value on the Lie algebra
$\mathfrak{u}(N)=T_{e}(U(N))$ is a scaled version of the Hilbert--Schmidt inner product:%
\begin{equation}
\left\langle X,Y\right\rangle =N\mathrm{trace}(X^{\ast}Y). \label{HSN}%
\end{equation}
We then express the Wilson loop functionals using the \textit{normalized}
trace,%
\[
\mathrm{tr}(A):=\frac{1}{N}\mathrm{trace}(A).
\]

\begin{figure}[ptb]%
\centering
\includegraphics[
height=2.1949in,
width=2.1949in
]%
{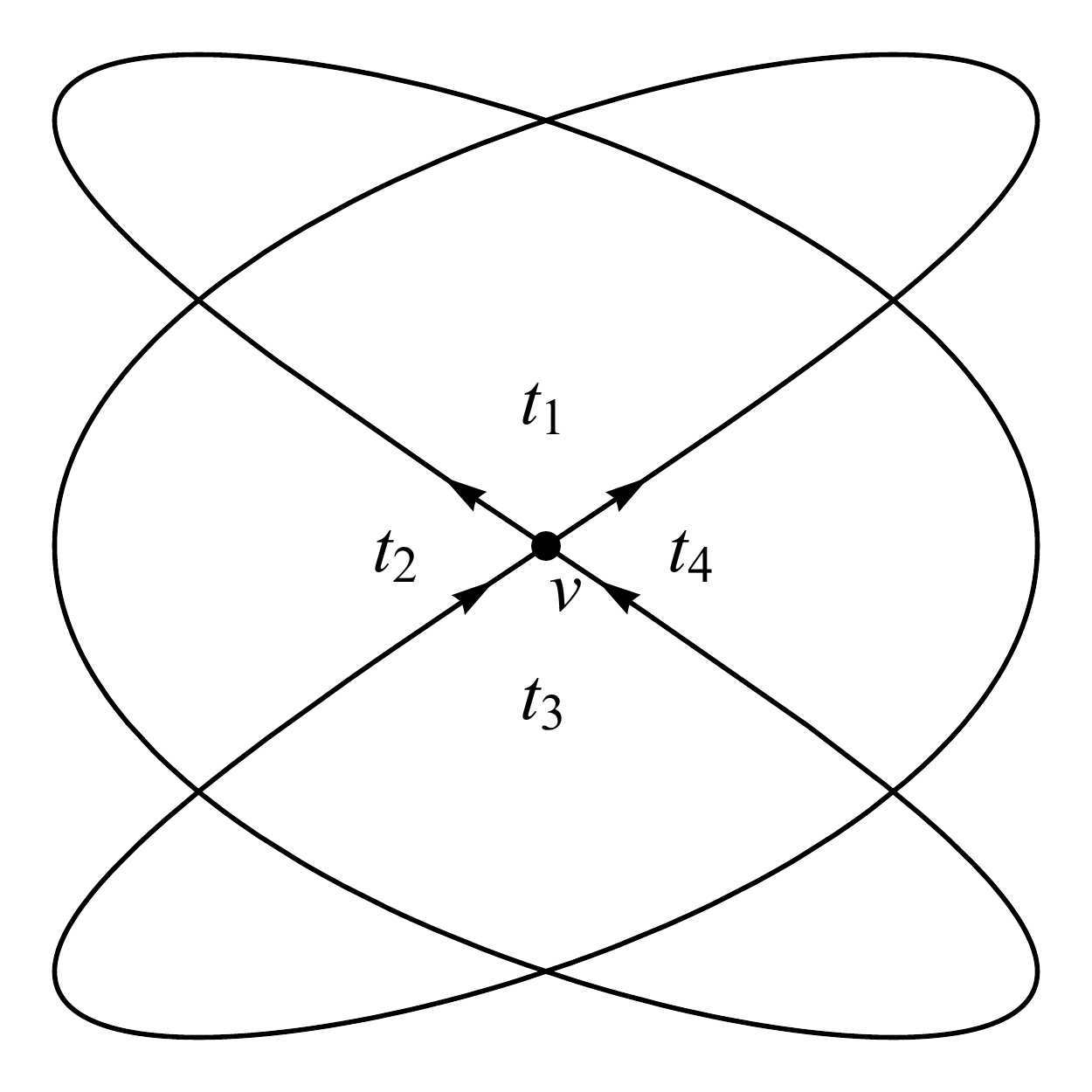}%
\caption{A typical loop $L$ for the Makeenko--Migdal equation}%
\label{mmplot.fig}%
\end{figure}

\begin{figure}[ptb]%
\centering
\includegraphics[
height=2.1949in,
width=2.1949in
]%
{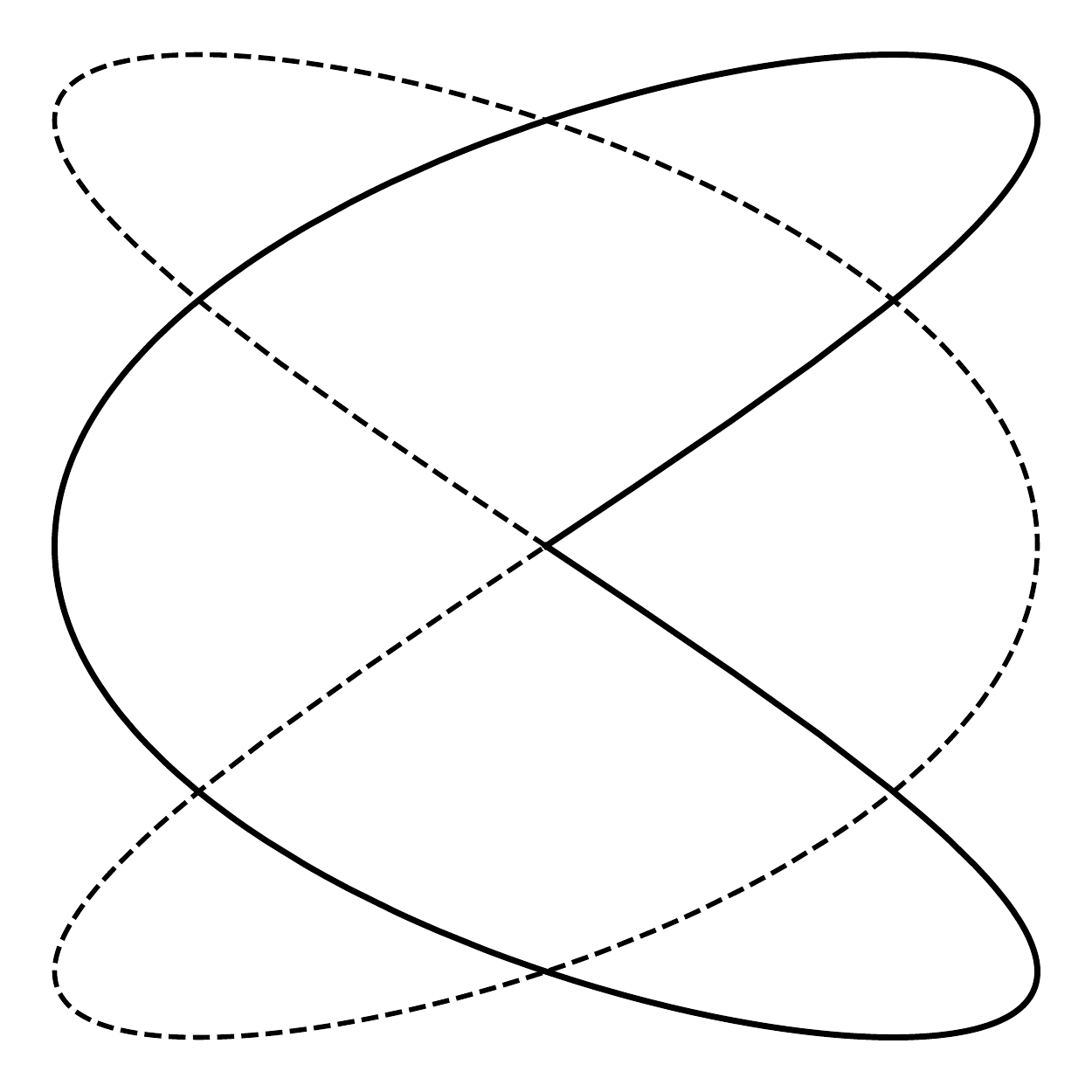}%
\caption{The loops $L_{1}$ (black) and $L_{2}$ (dashed)}%
\label{l1l2.fig}%
\end{figure}

We now consider a loop $L$ in the plane with simple crossings, and we let $v$
be one such crossing. We let $t_{1},$ $t_{2},$ $t_{3},$ and $t_{4}$ denote the
areas of the faces adjacent to $v,$ as in Figure \ref{mmplot.fig}. We also let
$L_{1}$ denote the portion of the loop from the beginning to the first return
to $v$ and let $L_{2}$ denote the loop from the first return to the end, as in
Figure \ref{l1l2.fig}. The planar Makeenko--Migdal equation then gives a
formula for the alternating sum of the derivatives of the Wilson loop
functional with respect to these areas:%
\begin{equation}
\left(  \frac{\partial}{\partial t_{1}}-\frac{\partial}{\partial t_{2}}%
+\frac{\partial}{\partial t_{3}}-\frac{\partial}{\partial t_{4}}\right)
\mathbb{E}\{\mathrm{tr}(\mathrm{hol}(L))\}=\mathbb{E}\{\mathrm{tr}%
(\mathrm{hol}(L_{1}))\mathrm{tr}(\mathrm{hol}(L_{2}))\}.\label{MMun}%
\end{equation}
We follow the convention that if any of the adjacent faces is the unbounded
face, the corresponding derivative on the left-hand side of (\ref{MMun}) is
omitted. Note also that the faces $F_{1},$ $F_{2},$ $F_{3},$ and $F_{4}$ are
not necessarily distinct, so that the same derivative may occur more than once
on the left-hand side of (\ref{MMun}).

The first rigorous proof of (\ref{MMun}) was given by L\'{e}vy in
\cite[Proposition 6.24]{LevyMaster}. A second proof was given by A. Dahlqvist
in \cite[Proposition 7.2]{Dahl}. Both of these proofs proceed by computing the
\textit{individual} time derivatives on the left-hand side of (\ref{MMun}).
These formulas involve calculations along a sequence of faces proceeding from
a face adjacent to $v$ to the unbounded face. After taking the alternating sum
of derivatives, both L\'{e}vy and Dahlqvist obtain a cancellation that allows
the result to simplify to the right-hand side of (\ref{MMun}). In \cite{DHK2},
three of the authors of the present paper gave three new proofs of
(\ref{MMun}). All of these proofs were \textquotedblleft
local\textquotedblright\ in nature, meaning that the calculations involve only
faces and edges adjacent to the crossing $v.$%

The goal of the present paper is to demonstrate that two of the proofs of
(\ref{MMun}) in \cite{DHK2} can be applied almost without change to the case
of an arbitrary compact surface $\Sigma$, possibly with boundary. In particular,
due to the local nature of the proofs in \cite{DHK2}, we do not require the presence
of an unbounded face.

Let us say that a graph $\mathbb{G}$ in $\Sigma$ is \textbf{admissible} if
$\mathbb{G}$ contains the entire boundary of $\Sigma$ and each component of
the complement of $\mathbb{G}$ is homeomorphic to a disk. (Actually, according
to Proposition 1.3.10 of \cite{LevyMarkov}, if each component of the
complement is a disk, the graph necessarily contains the entire boundary of
$\Sigma.$)

\begin{theorem}
[Makeenko--Migdal Equation for Surfaces]\label{mmSigma.thm}Let $\Sigma$ be a
compact surface, possibly with boundary. Let $K=U(N)$ and let $\mathbb{E}$
denote expectation value with respect to the normalized Yang--Mills measure
over $\Sigma,$ possibly with constraints on the holonomies around the boundary
components. Suppose that $L$ is a closed curve that can be traced out on an
admissible graph $\mathbb{G}$ in $\Sigma.$ Suppose $v$ is a simple crossing of
$L$ in the interior of $\Sigma$ and let $L_{1}$ and $L_{2}$ denote the two
pieces of the curve cut at $v.$ Then we have%
\[
\left(  \frac{\partial}{\partial t_{1}}-\frac{\partial}{\partial t_{2}}%
+\frac{\partial}{\partial t_{3}}-\frac{\partial}{\partial t_{4}}\right)
\mathbb{E}\{\mathrm{tr}(\mathrm{hol}(L))\}=\mathbb{E}\{\mathrm{tr}%
(\mathrm{hol}(L_{1}))\mathrm{tr}(\mathrm{hol}(L_{2}))\}.
\]

\end{theorem}

We will actually prove an abstract Makeenko--Migdal equation (generalizing Proposition
6.22 in \cite{LevyMaster}) that applies to an arbitrary structure group $K$
and that implies Theorem \ref{mmSigma.thm} as a special case. As in the plane
case, the abstract Makeenko--Migdal equation allows one to compute alternating
sums of derivatives of other sorts of functions; see Section 2.5 of
\cite{DHK2} for examples.

For any fixed $N,$ the Makeenko--Migdal equation in (\ref{MMun}) or in Theorem
\ref{mmSigma.thm} is not especially helpful in computing Wilson loop
functionals. After all, even though the loops $L_{1}$ and $L_{2}$ are simpler
than $L,$ the right-hand side of (\ref{MMun}) involves the \textit{expectation
of a product} of traces rather than a product of expectations. Thus, the
right-hand side cannot be considered as a recursively known quantity. In the
plane case, however, it is known that the Yang--Mills theory for $U(N)$ has a
large-$N$ limit, and that in this limit, all traces become deterministic.
(This deterministic limit is known as the master field and was investigated by
various authors, including I. M. Singer \cite{Sing}, R. Gopakumar and D. Gross
\cite{GG}, and M. Anshelevich and Sengupta \cite{AS}. A detailed proof of the
existence and deterministic nature of the limit was provided by L\'{e}vy in
\cite[Section 5]{LevyMaster}.) Thus, in the large-$N$ limit in the plane case,
there is no difference between the expectation of a product and a product of
expectations and (\ref{MMun}) becomes%
\begin{equation}
\left(  \frac{\partial}{\partial t_{1}}-\frac{\partial}{\partial t_{2}}%
+\frac{\partial}{\partial t_{3}}-\frac{\partial}{\partial t_{4}}\right)
\tau(\mathrm{hol}(L))=\tau(\mathrm{hol}(L_{1}))\tau(\mathrm{hol}(L_{2})),
\label{mmLargeN}%
\end{equation}
where $\tau(\cdot)$ is the limiting value of $\mathbb{E}\{\mathrm{tr}%
(\cdot)\}.$

In the plane case, L\'{e}vy also establishes the following \textquotedblleft
unbounded face condition.\textquotedblright\ If $t$ denotes the area of any
face $F$ that adjoins the unbounded face, we have%
\begin{equation}
\frac{\partial}{\partial t}\tau(\mathrm{hol}(L))=-\frac{1}{2}\tau
(\mathrm{hol}(L)),\quad(F\text{ adjoins the unbounded face}).
\label{simpleDeriv}%
\end{equation}
(See Axiom $\Phi_{4}$ in Section 0.5 of \cite{LevyMaster} and compare Theorem
2 in \cite{DHK2}.) L\'{e}vy then shows that the large-$N$ limit of $U(N)$
Yang--Mills theory on the plane is completely determined by the large-$N$
Makeenko--Migdal equation (\ref{mmLargeN}) and the unbounded face condition
(\ref{simpleDeriv}), together with some continuity and invariance properties
\cite[Section 0.5]{LevyMaster}.

It is currently not rigorously known whether Yang--Mills theory on a compact surface
$\Sigma$ admits a large-$N$ limit.  (However, see for example \cite{DaulKazakov}, where
the large-$N$ limit of Yang-Mills theory on the $2$-sphere is explored non-rigorously.) If the limit does exist and is
deterministic (as in the plane case), it is reasonable to expect that the
limiting theory would satisfy \eqref{mmLargeN} (this is assumed in \cite{DaulKazakov}).
One would have to justify interchanging the derivatives with the large-$N$ limit in Theorem
\ref{mmSigma.thm}. On the other hand, since $\Sigma$ does not have an
unbounded face, the unbounded face condition in (\ref{simpleDeriv}) does not
even make sense. Thus, even if (\ref{mmLargeN}) holds for the large-$N$ limit
of Yang--Mills theory on $\Sigma$, this relation may not allow for a complete
characterization of the limit. Nevertheless, if the large-$N$ limit on
$\Sigma$ exists and satisfies (\ref{mmLargeN}), this relation should contain a
lot of information about the limiting theory.

The authors thank Ambar Sengupta for many useful discussions of Yang--Mills
theory on surfaces.

\section{Yang--Mills theory on surfaces\label{YMsurf.sec}}

The Yang--Mills measure for a graph $\mathbb{G}$ in a surface $\Sigma$ has
been described by Sengupta, first for closed surfaces in \cite{Sen93} (see
also \cite{Sen97}) and then for surfaces with boundary \cite{Sen3}, possibly
incorporating constraints on the holonomy around the boundary. Related work
was done by Fine \cite{Fine1,Fine2} and Witten \cite{Witten1,Witten2}.
Sengupta's results were further developed and generalized by L\'{e}vy in
\cite{LevySurf} and \cite{LevyMarkov}.

We consider a compact, connected surface $\Sigma,$ possibly with boundary. We
do not require that $\Sigma$ be orientable. We then consider a connected
compact group $K,$ called the structure group, equipped with a fixed
bi-invariant Riemannian metric. (If $K$ is not simply connected, the
Yang--Mills measure as described below may incorporate contributions from
inequivalent principal $K$-bundles over $\Sigma$.) We also consider the heat
kernel $\rho_{t}$ on $K$ at the identity, that is, the unique function such
that%
\[
\frac{\partial\rho_{t}}{\partial t}=\frac{1}{2}\Delta\rho_{t}%
\]
and such that for any continuous function $f$ on $K,$%
\[
\lim_{t\rightarrow0}\int_{K}f(x)\rho_{t}(x)~dx=f(\mathrm{id}),
\]
where $\mathrm{id}$ is the identity element of $K$ and $dx$ is the normalized
Haar measure.

\subsection{The unconstrained Yang--Mills measure on a graph}

We begin by precisely defining the appropriate notion of a graph in $\Sigma$. By an edge we will
mean a continuous map $\gamma:[0,1]\rightarrow\Sigma$, assumed to be
injective except possibly that $\gamma(0)=\gamma(1).$ We identify two edges if
they differ by an orientation-preserving reparametrization. Two edges that
differ by an orientation-reversing reparametrization are said to be inverses
of each other. A graph is then a finite collection of edges (and their inverses) that meet only at
their endpoints. Given a graph $\mathbb{G},$ we choose arbitrarily one element
out of each pair consisting of an edge and its inverse. We then refer to the
chosen edges as the positively oriented edges.

We call a graph $\mathbb{G}$ in $\Sigma$ \textbf{admissible} if $\mathbb{G}$
contains the entire boundary of $\Sigma$ and each face $F$ of $\mathbb{G}$
(i.e., each component of the complement of $\mathbb{G}$ in $\Sigma$) is
homeomorphic to an open disk. Thus, the boundary of $F$ can be represented by
a single loop in $\mathbb{G}.$ To each positively oriented edge $e$ in $\mathbb{G}$
we associate an {\em edge variable} $x\in K$, and then correspondingly associate $x^{-1}$ to the inverse
of $e$. We then form a measure on $K^{n}$,
where $n$ is the number of edges, as follows. For each face $F$ of
$\mathbb{G}$, we consider the \textquotedblleft holonomy\textquotedblright%
\ $h$, which is just the product of edge variables (and their inverses) along the boundary of $F$.
We then consider first an un-normalized measure $\tilde{\mu}$ on $K^{n}$,
given by%
\[
d\tilde{\mu}(\mathbf{x})=\left(  \prod_{i}\rho_{\left\vert F_{i}\right\vert
}(h_{i})\right)  d\mathbf{x},
\]
where $d\mathbf{x}$ is the product of the \textit{normalized }Haar measures in
the edge variables.  Note: since the Haar measure on $K$ is symmetric (i.e.\ invariant
under $x\mapsto x^{-1}$), the measure $\tilde{\mu}$ is independent of the choice of
which edges in $\mathbb{G}$ are positively oriented.

We consider also the normalized measure%
\[
d\mu(\mathbf{x})=\frac{1}{Z}d\tilde{\mu}(\mathbf{x}),
\]
where%
\[
Z=\int_{K^{n}}\left(  \prod_{i}\rho_{\left\vert F_{i}\right\vert }%
(h_{i})\right)  d\mathbf{x}%
\]
is the \textbf{partition function} of the graph. This formula for $\mu$ is
\textit{Sengupta's formula} \cite[Theorem 5.3]{Sen93}, which he derives from a
rigorous version of the usual path-integral formula.  (As with $\tilde{\mu}$, $\mu$
is independent of which edges are chosen to be positively oriented.) We use the notation
$\mathbb{E}$ for the expectation value with respect to the normalized
Yang--Mills measure:%
\[
\mathbb{E}\{f\}:=\int_{K^{n}}f(\mathbf{x})~d\mu(\mathbf{x}).
\]

It is known that the partition function $Z$ depends only on the area and
diffeomorphism class of $\Sigma$ and not on the choice of graph; see
Proposition 5.2 in \cite{Sen93}. (For the independence of $Z$ from the graph, it
is essential that we use normalized Haar measures in the definition of the
un-normalized measure $\tilde{\mu}.$) If, for example, $\Sigma=S^{2},$ then
$Z$ is given by%
\[
Z_{S^{2}}=\rho_{A}(\mathrm{id}),
\]
where $A$ is the area of the sphere and $\mathrm{id}$ is the identity element
of $K.$ In particular, for a fixed diffeomorphism class of surface and fixed
topological type of the embedded graph, $Z$ depends only on the \textit{sum}
of the areas $t_{i}$ of the faces of $\mathbb{G}.$

Although the formula for the Yang--Mills measure on a surface is similar to
the formula \cite[Theorem 6.4]{Dr} in the plane case, the two measures behave
differently. In the plane case, the holonomies $h_{i}$ around the bounded
faces of a graph are independent heat-kernel distributed random variables
\cite[Proposition 4.4]{LevyMaster}. For a general compact surface $\Sigma,$
the $h_{i}$'s are neither independent nor heat kernel distributed. For the
case of a simple closed curve in $S^{2},$ for example, we may represent the
curve by a graph with a single edge, with edge variable $x.$ The holonomies
associated to the two faces of the graph are then $h_{1}=x$ and $h_{2}%
=x^{-1},$ so that the Yang--Mills measure for this graph is%
\[
d\mu(x)=\frac{1}{\rho_{s+t}(\mathrm{id})}\rho_{s}(x)\rho_{t}(x^{-1})~dx,
\]
where $s$ and $t$ are the areas of the two faces. (This formula may be
interpreted as saying that the holonomy around the loop is distributed as a
Brownian bridge at time $s,$ where the bridge returns to the identity at time
$s+t$.) Even for this simple example, there is no easy way to compute the
expected trace of the holonomy around the loop.

\begin{figure}[ptb]%
\centering
\includegraphics[
height=2.2139in,
width=2.5564in
]%
{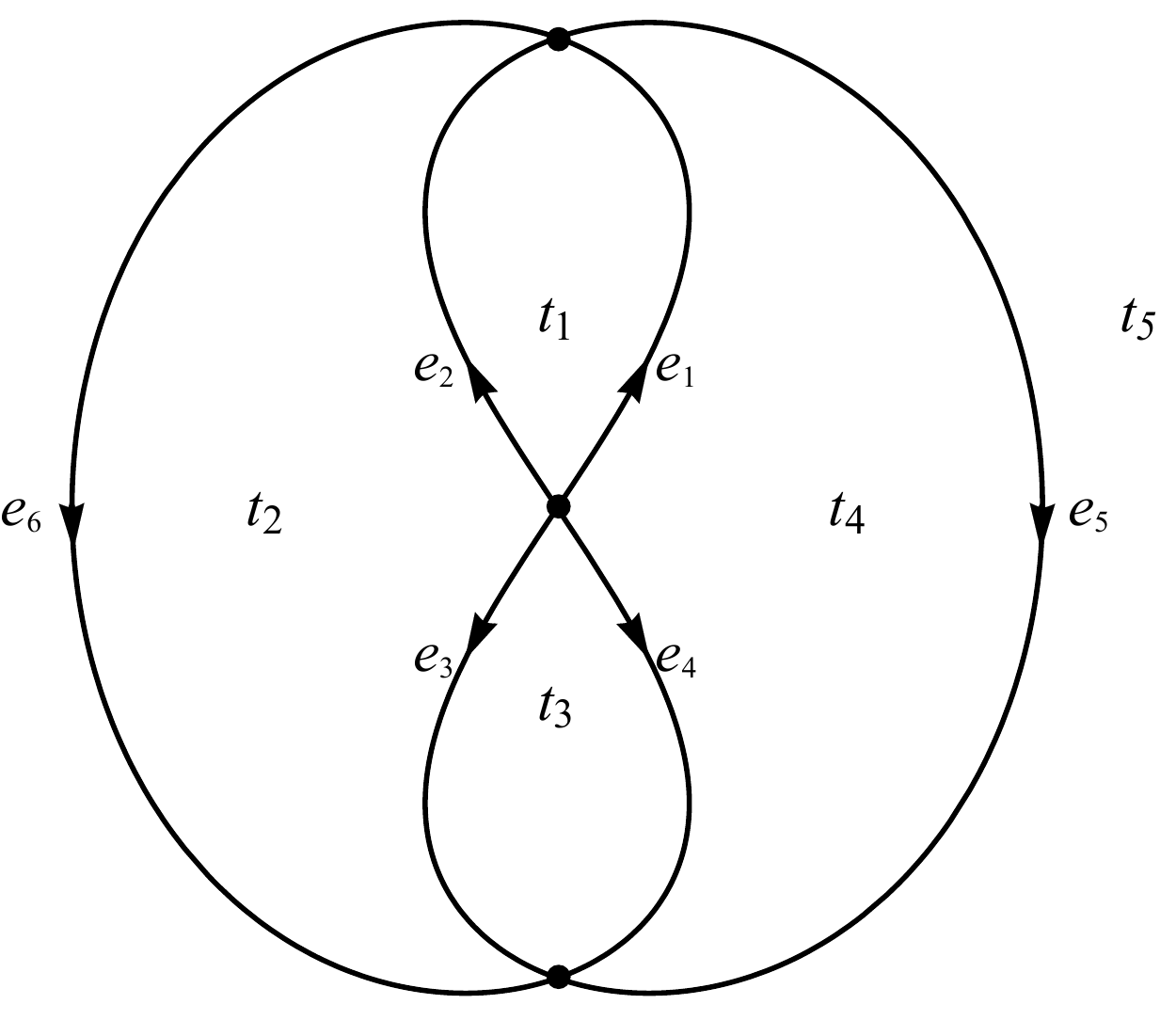}%
\caption{A graph embedded in $S^{2}$ with five faces}%
\label{surfexample.fig}%
\end{figure}

Although the Yang--Mills measure on a surface is more difficult to compute
with than the measure on the plane, we will show that two of the proofs of the
Makeenko--Migdal equation given in \cite{DHK2} go through essentially without
change. To illustrate this point, consider the graph in Figure
\ref{surfexample.fig}, which we regard as being embedded in $S^{2}.$ If
$x_{i}$ is the edge variable associated to the edge $e_{i},$ and $t_{j}$ is
the area of $F_{j},$ the un-normalized Yang--Mills measure takes the form%
\begin{equation}
d\tilde{\mu}(\mathbf{x})=\rho_{t_{1}}(x_{2}^{-1}x_{1})\rho_{t_{2}}(x_{3}%
^{-1}x_{6}x_{2})\rho_{t_{3}}(x_{4}^{-1}x_{3})\rho_{t_{4}}(x_{1}^{-1}x_{5}%
^{-1}x_{4})\rho_{t_{5}}(x_{6}^{-1}x_{5})~d\mathbf{x}. \label{muExample}%
\end{equation}
(Note that the boundary of, say, $F_{1}$ is $e_{1}e_{2}^{-1},$ but since
parallel transport is order-reversing, the holonomy around $F_{1}$ is
represented as $x_{2}^{-1}x_{1}.$)

If the graph were embedded in the plane instead of the sphere, we would simply
omit the factor of $\rho_{t_{5}}(x_{6}^{-1}x_{5}),$ since in that case,
$F_{5}$ would be the unbounded face, which does not contribute to Driver's
formula. We see, then, that replacing the plane by some other surface does not
change the \textquotedblleft local\textquotedblright\ structure of the
un-normalized Yang--Mills measure. If, for example, we wish to establish the
Makeenko--Migdal equation for the central vertex in Figure
\ref{surfexample.fig}, the first two proofs in \cite{DHK2} apply without
change, since the \textquotedblleft local Makeenko--Migdal
equation\textquotedblright\ in Theorem 6 there can be applied to the
integration over the variables $x_{1},\ldots,x_{4}.$  (In particular, since our proofs
in the plane case make no reference to the unbounded face, the absence of an
unbounded face on $\Sigma$ does not cause a difficulty.) Once the Makeenko--Migdal
equation for the un-normalized measure is established, it is then a simple
matter to establish it for the normalized measure as well.

\subsection{The constrained Yang--Mills measure on a graph}

It is possible to modify the construction in the preceding subsection by
constraining the holonomy around one or more of the boundary components to lie
in a fixed conjugacy class. If the boundary component in question consists of
a sequence $e_{1},\ldots,e_{k}$ of edges with edge variables $x_{1}%
,\ldots,x_{k},$ the holonomy around the component will be $x_{k}x_{k-1}\cdots
x_{1}$, since holonomy is order reversing. (Note that this boundary component
will usually not be the boundary of one of the faces of $\mathbb{G}.$) To
constrain $x_{k}x_{k-1}\cdots x_{1}$ to lie in $C,$ we insert a $\delta
$-function $\delta(x_{k}x_{k-1}\cdots x_{1}c^{-1})$ and then integrate over
$c\in C.$ Thus, integration with respect to the un-normalized constrained
measure $\tilde{\mu}$ takes the form%
\begin{align}
\int f(\mathbf{x})~d\tilde{\mu}(\mathbf{x})=  &  \int_{K^{n}}\int_{C_{1}%
}\cdots\int_{C_{N}}f(\mathbf{x})\left(  \prod_{i}\rho_{\left\vert
F_{i}\right\vert }(h_{i})\right) \nonumber\\
&  \times\prod_{j}\delta(x_{k_{j}}^{j}x_{k_{j}-1}^{j}\cdots x_{1}^{j}%
c_{j}^{-1})~d\mathbf{x}~d\mathrm{vol}(c_{1})~\cdots~d\mathrm{vol}(c_{N}),
\label{constrainedMu}%
\end{align}
where $C_{1},\ldots,C_{N}$ are the conjugacy classes to which various boundary
holonomies are constrained and where $d\mathrm{vol}$ is the normalized,
Ad-invariant volume measure on the given conjugacy class. (See Theorem 4 in
\cite{Sen3} and compare Section 1.5 of \cite{LevySurf}.) In
(\ref{constrainedMu}), we may interpret $\delta(\cdot)$ as the
small-$\varepsilon$ limit of $\rho_{\varepsilon}(\cdot).$ Alternatively, we
may think of $\delta(x_{k}x_{k-1}\cdots x_{1}c^{-1})$ as a rule telling
us that instead of integrating over $x_{k},$ we simply evaluate $x_{k}$ to
$(x_{k-1}\cdots x_{1}c^{-1})^{-1}.$

We may then construct a normalized measure by dividing by a normalization
constant, which we refer to as the constrained partition function. Similarly
to the unconstrained case, the constrained partition function depends only on
the area, the topological type of the surface, and the constraints, but not on
the choice of graph. (See the formula for $N_{T}(\mathbf{c})$ in Theorem 4 of
\cite{Sen3} and compare Proposition 4.3.5 in \cite{LevyMarkov} in a more
general setting.)

\begin{figure}[ptb]%
\centering
\includegraphics[
height=2.5564in,
width=2.5564in
]%
{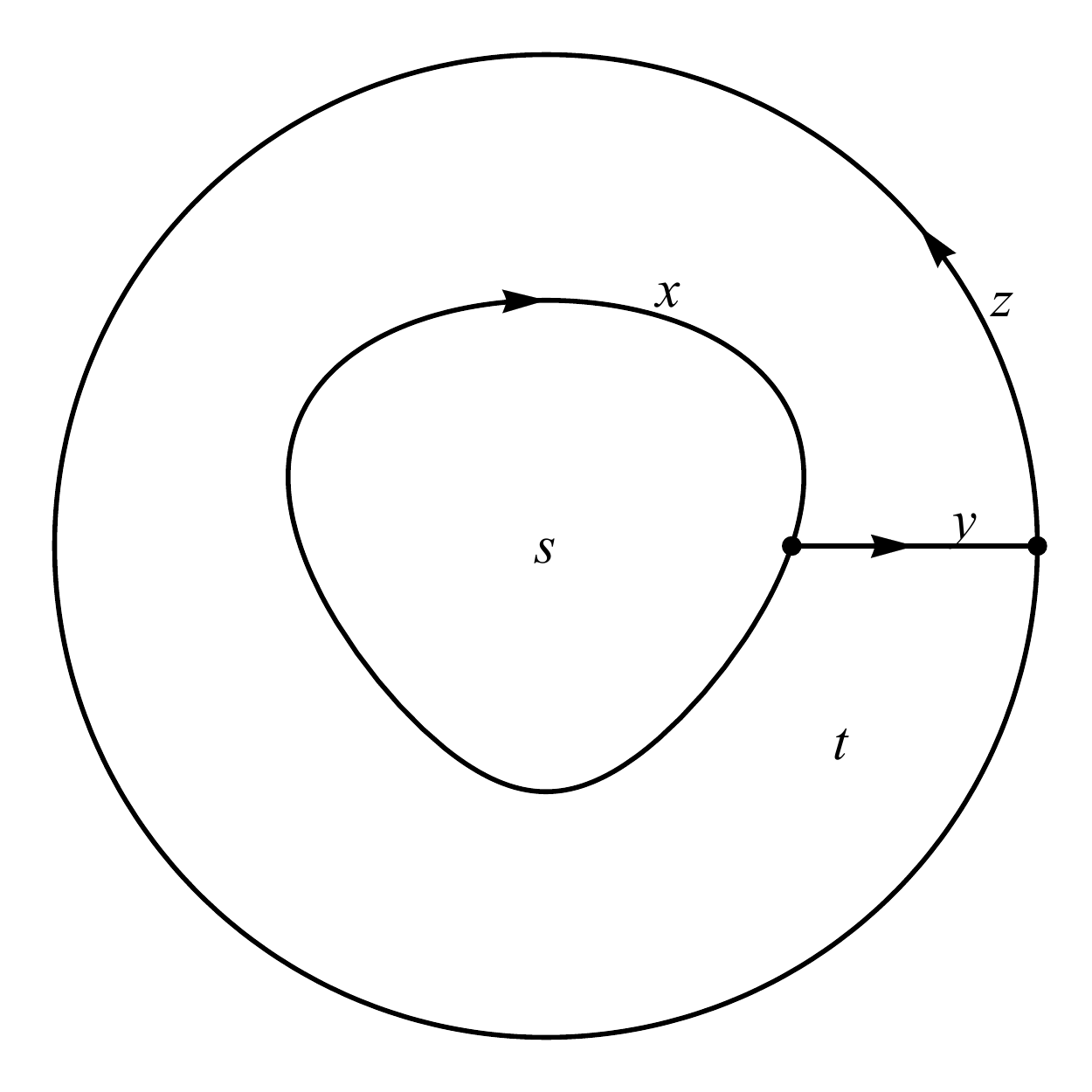}%
\caption{The holonomy $z$ around the boundary of the disk is constrained to
lie in $C.$}%
\label{constrainedintegral.fig}%
\end{figure}

In Figure \ref{constrainedintegral.fig}, for example, if the holonomy around
the boundary of a disk is constrained to lie in $C$, the expected trace of the
holonomy around the inner loop would be computed as%
\begin{align*}
\frac{1}{Z}  &  \int_{C}\int_{K^{3}}\mathrm{tr}(x^{-1})\rho_{s}(x^{-1}%
)\rho_{t}(y^{-1}zyx)\delta(zc^{-1})~dx~dy~dz~d\mathrm{vol}(c)\\
&  =\frac{1}{Z}\int_{C}\int_{K^{2}}\mathrm{tr}(x^{-1})\rho_{s}(x^{-1})\rho
_{t}(y^{-1}cyx)~dx~dy~d\mathrm{vol}(c)\\
&  =\frac{1}{Z}\int_{C}\int_{K}\mathrm{tr}(x^{-1})\rho_{s}(x^{-1})\rho
_{t}(cx)~dx~d\mathrm{vol}(c),
\end{align*}
where in the last expression, we have used the Ad-invariance of $d\mathrm{vol}%
(c)$ to eliminate the $y$ variable. By contrast, if we left the boundary
holonomy unconstrained, we would integrate $\mathrm{tr}(x^{-1})\rho_{s}%
(x^{-1})\rho_{t}(y^{-1}zyx)$ over $K^{3}$, in which case the result would
simplify to $\int_{K}\mathrm{tr}(x^{-1})\rho_{s}(x^{-1})~dx$ (with no
normalization factor necessary).%

\section{The Makeenko--Migdal equation for surfaces}

Throughout this section, we assume $\mathbb{G}$ is an admissible graph in
$\Sigma,$ that is, one containing the boundary of $\Sigma$ and such that each
face of $\mathbb{G}$ is a disk.

\subsection{An abstract Makeenko--Migdal equation}

Following L\'{e}vy \cite[Definition 6.21]{LevyMaster} for the plane case, we
now introduce a natural invariance property that will be crucial in proving
the Makeenko--Migdal equation.

\begin{definition}
\label{extendedInv.def}Let $\mathbb{G}$ be an admissible graph in $\Sigma$ and
let $v$ be a vertex of $\mathbb{G}$ in the interior of $\Sigma$ having exactly
four distinct edges, labeled in cyclic order as $e_{1},e_{2},e_{3},e_{4}$ and
taken to be outgoing edges. Let $f:K^{n}\rightarrow\mathbb{C}$ be a function
of the edge variables of $\mathbb{G}$ and let $a_{1},a_{2},a_{3},a_{4}$ be the
edge variables associated to $e_{1},e_{2},e_{3},e_{4}.$ Then $f$ has
\textbf{extended gauge invariance} at $v$ if%
\[
f(a_{1},a_{2},a_{3},a_{4},\mathbf{b})=f(a_{1}x,a_{2},a_{3}x,a_{4}%
,\mathbf{b})=f(a_{1},a_{2}x,a_{3},a_{4}x,\mathbf{b})
\]
for all $x\in K,$ where $\mathbf{b}$ is the tuple of all edge variables other
than $a_{1},a_{2},a_{3},a_{4}.$
\end{definition}

With this definition in hand, we may formulate a general version of the
Makeenko--Migdal equation for $\Sigma,$ generalizing Proposition 6.22 in
\cite{LevyMaster} in the plane case. The result applies to arbitrary structure
groups $K$ and to functions that are not necessarily given as the trace of a
holonomy. In what follows, we allow the areas of the faces to be arbitrary
positive real numbers; if we vary one area with the other areas fixed, we are
changing the total area of the surface.

\begin{figure}[ptb]%
\centering
\includegraphics[
height=2.25in,
width=2.25in
]%
{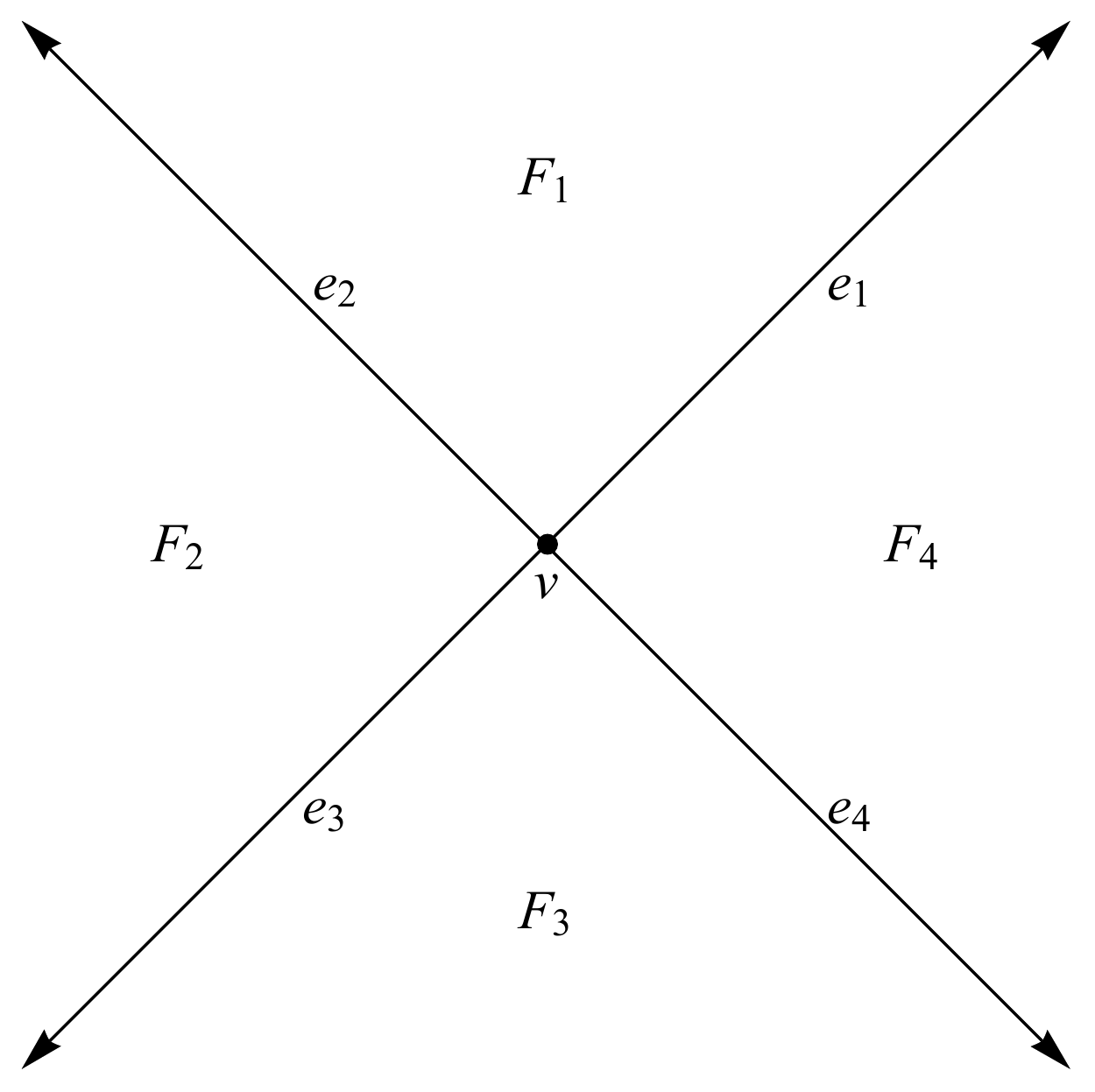}%
\caption{The edges and faces at $v$}%
\label{edgesandfaces.fig}%
\end{figure}

We consider a graph with four distinct edges $e_{1},\ldots,e_{4}$ attached to a
vertex $v,$ and we label the four faces surrounding $v$ as $F_{1},\ldots
,F_{4},$ as in Figure \ref{edgesandfaces.fig}, with the labeling chosen so
that $e_{1}$ lies between $F_{4}$ and $F_{1}.$%

\begin{theorem}
[Abstract Makeenko--Migdal Equation for $\Sigma$]\label{abstractMM.thm}Following the
notation of Definition \ref{extendedInv.def}, assume the four faces
and four edges adjacent to $v$ are distinct. Suppose $f:K^{n}\rightarrow
\mathbb{C}$ is a smooth function with extended gauge invariance at $v$. If
$t_{1},\ldots,t_{4}$ denote the areas of the faces of $\mathbb{G}$ surrounding
$v,$ we have
\[
\left(  \frac{\partial}{\partial t_{1}}-\frac{\partial}{\partial t_{2}}%
+\frac{\partial}{\partial t_{3}}-\frac{\partial}{\partial t_{4}}\right)
\int_{K^{n}}f~d\mu=-\int_{K^{n}}\nabla^{a_{1}}\cdot\nabla^{a_{2}}f~d\mu,
\]
where $\mu$ is the normalized Yang--Mills measure, possibly with constraints
on the boundary holonomies. The same result holds with $\mu$ replaced by the
un-normalized measure $\tilde{\mu}.$
\end{theorem}

Using the arguments in Section 4 of \cite{DHK2}, it is possible to prove this
result also when the faces are not distinct. It is also possible to formulate
and prove a version of the result when the four edges emanating from $v$ are
not distinct, although the definition of extended gauge invariance needs some
modification in this case. See Section \ref{nongeneric.sec} for more information.

In the theorem, the gradients are left-invariant gradients with respect to
$a_{1}$ and $a_{2}$ with the other edge variables fixed. Explicitly,%
\[
(\nabla^{a_{1}}\cdot\nabla^{a_{2}}f)(a_{1},a_{2},a_{3},a_{4},\mathbf{b}%
)=\sum_{X}\left.  \frac{\partial^{2}}{\partial s\partial t}f(a_{1}e^{sX}%
,a_{2}e^{tX},a_{3},a_{4},\mathbf{b})\right\vert _{s=t=0},
\]
where $X$ ranges over an orthonormal basis for the Lie algebra $\mathfrak{k}$
of $K$ and $\mathbf{b}$ represents the tuple of edge variables other than
$a_{1},\ldots,a_{4}.$ Using the extended gauge invariance of $f,$ it is easy
to show that%
\[
\nabla^{a_{1}}\cdot\nabla^{a_{2}}f=-\nabla^{a_{2}}\cdot\nabla^{a_{3}}%
f=\nabla^{a_{3}}\cdot\nabla^{a_{4}}f=-\nabla^{a_{4}}\cdot\nabla^{a_{1}}f.
\]

Suppose $L$ is a closed curve traced out in $\mathbb{G}$ that has a crossing at
$v.$ Specifically, assume $L$ starts at $v,$ leaves $v$ along $e_{1},$ returns
to $v$ along $e_{4}^{-1},$ leaves $v$ again along $e_{2},$ and then finally
returns to $v$ along $e_{3}^{-1}$ (with no visits to $v$ besides those just
mentioned). Then since holonomy is order-reversing, we will have%
\[
\mathrm{tr}(\mathrm{hol}(L))=\mathrm{tr}(a_{3}^{-1}\alpha a_{2}a_{4}^{-1}\beta
a_{1}),
\]
where $\alpha$ and $\beta$ are words in the $\mathbf{b}$ variables. Any
function of the this form is easily seen to have extended gauge invariance. If
$K=U(N),$ we compute that%
\begin{align*}
\nabla^{a_{1}}\cdot\nabla^{a_{2}}[\mathrm{tr}(a_{3}^{-1}\alpha a_{2}a_{4}%
^{-1}\beta a_{1})]  &  =\sum_{X}\mathrm{tr}(a_{3}^{-1}\alpha a_{2}Xa_{4}%
^{-1}\beta a_{1}X)\\
&  =-\mathrm{tr}(a_{3}^{-1}\alpha a_{2})\mathrm{tr}(a_{4}^{-1}\beta a_{1})\\
&  =-\mathrm{tr}(L_{2})\mathrm{tr}(L_{1}),
\end{align*}
where $L_{1}$ and $L_{2}$ are as in Theorem \ref{mmSigma.thm}, and where we
used the elementary identity $\sum_{X}XCX=-\mathrm{tr}(C)I$ (e.g.,
\cite[Proposition 3.1]{DHK}) in the second equality. This calculation shows
that the abstract Makeenko--Migdal equation implies the Makeenko--Migdal
equation for $U(N)$ (Theorem \ref{mmSigma.thm}).

\subsection{The generic case}

Let us assume at first that our loop is traced out in an admissible graph
$\mathbb{G}$ and that the vertex $v$ is generic, meaning that the edges
$e_{1},\ldots,e_{4}$ and the faces $F_{1},\ldots,F_{4}$ are distinct. We then
make use of the following result, which was proven in \cite[Theorem 6]{DHK2}.

\begin{theorem}
[Local Abstract Makeenko--Migdal Equation]\label{localMM.thm}Suppose
$f:K^{4}\rightarrow\mathbb{C}$ is a smooth function satisfying the following
\textquotedblleft extended gauge invariance\textquotedblright\ property:%
\[
f(a_{1},a_{2},a_{3},a_{4})=f(a_{1}x,a_{2},a_{3}x,a_{4})=f(a_{1},a_{2}%
x,a_{3},a_{4}x)
\]
for all $\mathbf{a}=(a_{1},a_{2},a_{3},a_{4})$ in $K^{4}$ and all $x$ in $K.$
For each fixed $\mathbf{\alpha}=(\alpha_{1},\alpha_{2},\alpha_{3},\alpha_{4})$
in $K^{4}$ and $\mathbf{t}=(t_{1},t_{2},t_{3},t_{4})$ in $(\mathbb{R}^{+}%
)^{4},$ define a measure $\mu_{\mathbf{\alpha},\mathbf{t}}$ on $K^{4}$ by%
\[
d\mu_{\mathbf{\alpha},\mathbf{t}}(\mathbf{a)=}\rho_{t_{1}}(a_{2}^{-1}%
\alpha_{1}a_{1})\rho_{t_{2}}(a_{3}^{-1}\alpha_{2}a_{2})\rho_{t_{3}}(a_{4}%
^{-1}\alpha_{3}a_{3})\rho_{t_{4}}(a_{1}^{-1}\alpha_{4}a_{4})~d\mathbf{a},
\]
where $d\mathbf{a}$ is the normalized Haar measure on $K^{4}.$ Then for all
$\mathbf{\alpha}\in K^{4},$ we have
\[
\left(  \frac{\partial}{\partial t_{1}}-\frac{\partial}{\partial t_{2}}%
+\frac{\partial}{\partial t_{3}}-\frac{\partial}{\partial t_{4}}\right)
\int_{K^{4}}f~d\mu_{\mathbf{\alpha},\mathbf{t}}=-\int_{K^{4}}\nabla^{a_{1}%
}\cdot\nabla^{a_{2}}f~d\mu_{\mathbf{\alpha},\mathbf{t}}.
\]

\end{theorem}

We now come to the proof of the abstract Makeenko--Migdal equation in Theorem
\ref{abstractMM.thm}, in the generic case where the four edges $e_{1}%
,\ldots,e_{4}$ and the four faces $F_{1},\ldots,F_{4}$ are distinct.

\medskip

\begin{proof}
[of Theorem \ref{abstractMM.thm}]Let $\mathbf{b}$ denote the tuple of
all edge variables other than $a_{1},\ldots,a_{4}.$ The holonomies around the
adjoining faces $F_{i},$ $i=1,\ldots,4,$ will have the form%
\[
h_{i}=a_{i+1}^{-1}\alpha_{i}a_{i},
\]
where $\alpha_{i}$ is a word in the $\mathbf{b}$ variables. Let us first
consider integration with respect to the un-normalized Yang--Mills measure
$\tilde{\mu}$, with or without constraints on boundary holonomies. Since $v$
lies in the interior of $\Sigma,$ the edges $e_{1},\ldots,e_{4}$ do not lie on
the boundary. Thus, the holonomy around any boundary component will involve
only the $\mathbf{b}$ variables. Integration with respect to $\tilde{\mu}$
therefore takes the form of integration over $a_{1},\ldots,a_{4}$ with respect
to $\mu_{\mathbf{\alpha},\mathbf{t}},$ where $\mathbf{\alpha}$ is a function
of the $\mathbf{b}$ variables, followed by integration in the $\mathbf{b}$
variables and possibly another layer of integration with respect to the
constraint variables $c_{j}.$ In the un-normalized measure $\tilde{\mu},$ the
only dependence on $t_{1},\ldots,t_{4}$ is in the inner layer of integration.
Thus, we may push the time derivatives inside the outer layers of integration
and allow them to hit on the integral over~$K^{4}.$ If we then apply the local
result in Theorem \ref{localMM.thm}, Theorem \ref{abstractMM.thm} for
$\tilde{\mu}$ will follow.

For the normalized measure, we must incorporate the partition function $Z.$
Since $Z$ depends only on the total area of the surface (i.e., the sum the
areas of all the faces), we see that%
\[
\left(  \frac{\partial}{\partial t_{1}}-\frac{\partial}{\partial t_{2}}%
+\frac{\partial}{\partial t_{3}}-\frac{\partial}{\partial t_{4}}\right)  Z=0.
\]
Thus, Theorem \ref{abstractMM.thm} for the normalized measure easily follows
from the corresponding result for the un-normalized measure.  \hfill $\square$
\end{proof}

\subsection{The nongeneric case\label{nongeneric.sec}}%

\begin{figure}[ptb]%
\centering
\includegraphics[
height=1.5in,
width=3.1in
]%
{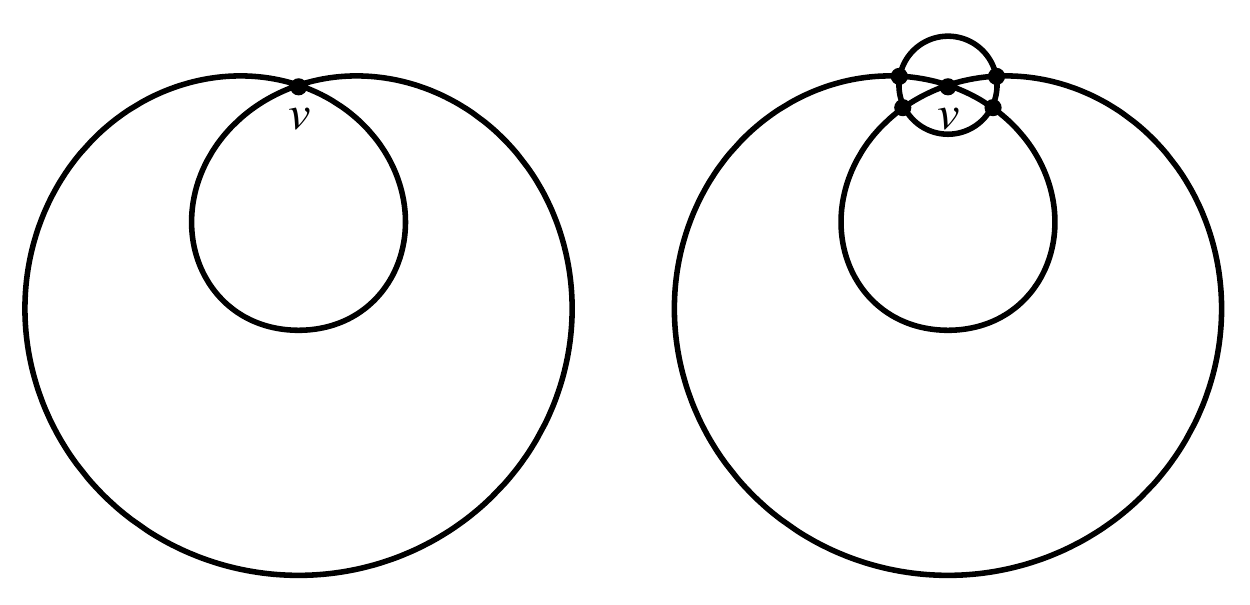}%
\caption{A graph that is non-generic at $v$ (left) and its generic
counterpart}%
\label{generic1.fig}%
\end{figure}

Suppose $\mathbb{G}$ is an admissible graph and $v$ is a vertex of
$\mathbb{G}$ having four attached edges, where we count an edge twice if both
ends of the edge are attached to $v.$ We say that $\mathbb{G}$ is
\textbf{nongeneric} at $v$ if either the four edges are not distinct or the
four faces surrounding $v$ are not distinct. If $\mathbb{G}$ is not generic at
$v,$ we can embed $\mathbb{G}$ into another admissible graph $\mathbb{G}%
^{\prime}$ that is generic at $v,$ as in Figure \ref{generic1.fig}. If $L$ is
a loop traced out on $\mathbb{G}$ with a simple crossing at $v,$ then
\textquotedblleft the same\textquotedblright\ loop can also be traced out on
$\mathbb{G}^{\prime}.$ In that case, the expectation values of $\mathrm{tr}%
(\mathrm{hol}(L))$ and of $\mathrm{tr}(\mathrm{hol}(L_{1}))\mathrm{tr}%
(\mathrm{hol}(L_{2}))$---where $L_{1}$ and $L_{2}$ are as in Theorem
\ref{mmSigma.thm}---are the same whether we work over $\mathbb{G}$ or over
$\mathbb{G}^{\prime}$. This invariance result has two parts. First, there is
invariance under subdividing an edge by adding a vertex in the middle of that
edge, which is very easy to establish, as shown in Section 4.1 of \cite{DHK2}.
(The argument given there applies equally well in the surface case or the
plane case.) Second, there is invariance under keeping the vertex set the same
and adding a new edge. This invariance result is a consequence of Proposition
4.3.4 in \cite{LevyMarkov}, in the case that the L\'{e}vy process there is
taken to be Brownian motion on $K.$

Furthermore, it is not hard to see that area derivatives of expectation values
give the same result whether computed over $\mathbb{G}$ or $\mathbb{G}%
^{\prime}.$ (See Section 4.3 of \cite{DHK2}.) Thus, the $U(N)$ version of the
Makeenko--Migdal equation for the loop in $\mathbb{G}$ reduces to the
corresponding result for the loop in $\mathbb{G}^{\prime},$ which in turn
follows from Theorem \ref{abstractMM.thm}. In the graph on the left-hand side
of Figure \ref{generic1.fig}, for example, $F_{1}$ coincides with $F_{3}.$
Thus, $t_{3}$ is just another name for $t_{1}$ and the Makeenko--Migdal
equation may be written as%
\[
\left(  2\frac{\partial}{\partial t_{1}}-\frac{\partial}{\partial t_{2}}%
-\frac{\partial}{\partial t_{4}}\right)  \mathbb{E}\{\mathrm{tr}%
(\mathrm{hol}(L))\}=\mathbb{E}\{\mathrm{tr}(\mathrm{hol}(L_{1}))\mathrm{tr}%
(\mathrm{hol}(L_{2}))\}.
\]

It is also possible to formulate a version of Theorem \ref{abstractMM.thm}
itself that holds in the nongeneric situation. If the edges $e_{1}%
,\ldots,e_{4}$ are distinct but the faces $F_{1},\ldots,F_{4}$ are not
distinct, Theorem \ref{abstractMM.thm} holds with no changes to the statement,
and the arguments in Section 4 of \cite{DHK2} show how this result can be
reduced to the generic case. If the edges (and possibly also faces) are not
distinct, the notion of extended gauge invariance needs some revision
\cite[Section 4.2]{DHK2}, after which one can reduce the result to the generic
case. Since this process of reduction requires no changes from the arguments
in \cite{DHK2}, we do not enter into the details here, but refer the
interested reader to Sections 4.2 and 4.3 of \cite{DHK2}.

\end{document}